\documentclass[aps,prl,twocolumn,superscriptaddress,showpacs]{revtex4-1}
\usepackage{amsmath}
\usepackage{mathtools}
\usepackage{amsfonts}
\usepackage{graphicx}
\usepackage{braket}
\usepackage{color}
\usepackage[T1]{fontenc}
\begin{document}

%\preprint{}

\title{Quantum correlations and complementarity of vectorial light fields}


\author{Andreas Norrman}
\email[]{andreas.norrman@mpl.mpg.de}
\affiliation{Max Planck Institute for the Science of Light, Staudtstra{\ss}e 2, D-91058 Erlangen, Germany}
\author{{\L}ukasz Rudnicki}
\affiliation{Max Planck Institute for the Science of Light, Staudtstra{\ss}e 2, D-91058 Erlangen, Germany}
\affiliation{Center for Theoretical Physics, Polish Academy of Sciences, Aleja Lotnik\'{o}w 32/46, 02-668 Warsaw, Poland}


\date{\today}

\begin{abstract}

We explore quantum correlations of general vector-light fields in multislit interference and show that the $n$th-order field-coherence matrix is directly linked with the reduced $n$-photon density matrix. The connection is utilized to examine photon wave-particle duality in the double-slit configuration, revealing that there is a hidden information-theoretic contribution that complements the standard inequality associated with such duality by transforming it into a strict equality, a triality identity. We also establish a general quantum complementarity relation among the field correlations and the particle correlations which holds for any number of slits, correlation orders, and vector-light states. The framework that we advance hence uncovers fundamental physics about quantum interference.

\end{abstract}

%\pacs{XXX, YYY, ZZZ}
%\keywords{}

\maketitle


\emph{Introduction}.---The seminal quantum theory of optical coherence \cite{Glauber63,Mandelbook}, dealing with field correlations of light, is ubiquitous in the physical sciences; it is widely exploited in quantum optics \cite{Fortsch13,Tenne19}, atomic physics \cite{Hodgman11,Perrin12}, optomechanics \cite{Cohen15,Ockeloen18}, quantum simulation \cite{Barrett13}, quantum electronics \cite{Brange15}, and cosmology \cite{Giovannini11}, among other research areas. Recently quantum coherence of genuine vector-light fields was examined in double-slit interference, revealing a new fundamental aspect of photon wave-particle duality \cite{Norrman17}. The rapid progress in quantum information science has at the same time led to an ever-growing interest towards nonclassical correlations that may prevail in multipartite quantum compositions of diverse physical nature \cite{Horodecki09,Chiara18}. These correlations lie at the heart of foundational quantum physics, with applications in quantum teleportation, quantum cryptography, and quantum computation. 

Quantum correlations of indistinguishable systems, in particular, have attracted broad interest lately \cite{Eckert02,Tichy11}. Such correlations may occur among modes \cite{Huang94,Zanardi01,Zanardi04}, usually met in operational quantum science, or between particles \cite{Paskauskas01,Li01,Ghirardi04}, mainly considered in many-body physics. The latter are regarded more fundamental as they are a prerequisite for the former \cite{Wiseman03}. In atomic physics, where correlation functions render schemes for joint probability detection of atoms \cite{Bach04}, Glauber's contribution \cite{Glauber63,Glauber99} can be viewed as a unified framework that subsumes field and particle correlations \cite{KRZ}. In quantum electrodynamics, however, where photons have no well-defined position operator \cite{IBB09}, particle-like information-theoretic content of the electromagnetic field is customarily omitted.

In this Letter, we investigate the relationship between \textit{field correlations} and \textit{particle correlations} of true \textit{vectorial light} of any quantum state in multislit interference. We show that the $n$th-order field correlations are directly connected to the particle correlations among $n$ photons. This relationship is especially employed to explore \textit{quantum complementarity} in the celebrated double-slit setup, resulting in the discovery of a tight equality which may be interpreted as describing photon \textit{wave-particle triality}. In the general case with any number of slits, we derive a fundamental complementarity relation between the $n$th-order field correlations and the $n$th-order particle correlations, stating that there is a strict \textit{trade-off} between these two types of correlations for all quantum states of light. Our work may be viewed as the most general framework regarding quantum complementarity of field-particle correlations in vector-light interference, providing deeper insights into foundational quantum-interference physics.

\emph{Modes and particles}.---Let us first discuss the notions of modes and particles in the description of quantum light. As an example we consider a pure, unit-trace normalized, two-photon state $\hat\rho^{(2)}=\left|\Psi\right\rangle\left\langle\Psi\right|$ involving two orthogonal modes ($x$ and $y$).

In the \textit{second-quantized} (mode) representation,
\begin{subequations}
\begin{equation}
\label{mode}
\left|\Psi\right\rangle=C_{1}\left|20\right\rangle+C_{2}\left|02\right\rangle+C_{3}\left|11\right\rangle,
\end{equation}
where the Fock state $\left|N_{x}N_{y}\right\rangle$ is understood as ``mode $x$ contains $N_{x}$ photons and mode $y$ contains $N_{y}$ photons''. In the \textit{first-quantized} (particle) representation \cite{footnote1},
\begin{equation}
\label{photon}
\left|\Psi\right\rangle=C_{1}\left|xx\right\rangle+C_{2}\left|yy\right\rangle+C_{3}\frac{\left|xy\right\rangle+\left|yx\right\rangle}{\sqrt{2}},
\end{equation}
where $\left|ij\right\rangle$ is interpreted as ``the first photon is in mode $i$ and the second photon is in mode $j$''. The symmetric last term in Eq.~(\ref{photon}) follows from bosonic indistinguishability of photons. Equations~(\ref{mode}) and (\ref{photon}) describe exactly the same state, and the first-quantized (rescaled \cite{footnote2}) matrix elements $\rho^{(2)}_{i_{1}i_{2},j_{1}j_{2}}\coloneqq 2\bra{i_{1}i_{2}}\hat{\rho}^{(2)}\ket{j_{1}j_{2}}$ are connected to the second-quantized language via
\begin{equation}
\rho^{(2)}_{i_{1}i_{2},j_{1}j_{2}}=\bra{\Psi}\hat{a}_{j_{1}}^{\dagger}\hat{a}_{j_{2}}^{\dagger}
\hat{a}_{i_{1}}\hat{a}_{i_{2}}\ket{\Psi},
\end{equation}
in which $\hat{a}_{j_{\alpha}}^{\dagger}$ ($\hat{a}_{i_{\alpha}}$) with $\alpha\in\{1,2\}$ is the creation (annihilation) operator of the mode $j_{\alpha}\in\{x,y\}$ ($i_{\alpha}\in\{x,y\}$). This ``duality of pictures'' clearly extends beyond the case of two photons in two modes, being applicable for every multimode and multiphoton mixed state. \end{subequations}

Nevertheless, the two representations contain very distinct information about the \textit{correlations} in the system. The \textit{reduced one-mode state} associated with Eq.~(\ref{mode}) and given by $\mathrm{diag}\left(|C_1|^2,|C_2|^2,|C_3|^2\right)$ describes \textit{entanglement between the two modes}. However, the \textit{reduced one-photon state} corresponding to Eq.~(\ref{photon}), with matrix elements
\begin{equation}
\label{reduced2}
\varrho_{i_{1},j_{1}}^{(2;1)}\coloneqq \sum_{k\in\{x,y\}}\rho_{i_{1}k,j_{1}k}^{(2)}\equiv  \bra{\Psi}\hat{a}_{j_{1}}^{\dagger}
\hat{a}_{i_{1}}\ket{\Psi},
\end{equation}
characterizes \textit{quantum correlations between the two photons}, which owing to the bosonic indistinguishability cannot be operationally classified as pure entanglement \cite{Li01}. While unitary operations achievable by linear optics may alter the mode correlations \cite{Enk03,JETP}, the primordial particle (photon) correlations \cite{Wiseman03} remain invariant.

\emph{Quantum correlations in vector-light interference}.---The complete information about the $n$th-order quantum-field correlations of \textit{vectorial light} \cite{footnote3}, at $2n$ space-time points $x_{1}=(\mathbf{r}_{1},t_{1}),\ldots,x_{2n}=(\mathbf{r}_{2n},t_{2n})$, is encoded in the $n$th-order correlation (or coherence) matrix \cite{Glauber63,Mandelbook}
\begin{align}
\label{Gn}
\mathbf{G}^{(n)}(x_{1},\ldots,x_{2n})&=\mathrm{tr}[\hat{\rho}\hat{\mathbf{E}}^{(-)}(x_{1})\otimes\cdots\otimes\hat{\mathbf{E}}^{(-)}(x_{n}) \nonumber \\ 
& \otimes\hat{\mathbf{E}}^{(+)}(x_{n+1})\otimes\cdots\otimes\hat{\mathbf{E}}^{(+)}(x_{2n})].
\end{align}
Here $\hat{\mathbf{E}}^{(+)}(x)$ and $\hat{\mathbf{E}}^{(-)}(x)$ are the positive and negative frequency parts of the electric-field operator, while $\hat{\rho}$ is the density operator of the quantum state.

Let us consider a vector-light field in $L$-slit interference. The electric-field operator at slit $m\in\{1,\ldots,L\}$ reads
\begin{equation}
\label{Em}
\hat{\mathbf{E}}^{(+)}(x_{m})=E_{0}\sum_{s=1}^{2}\hat{a}_{ms}\mathbf{e}_{ms}
e^{i(\mathbf{k}\cdot\mathbf{r}_{m}-\omega t_{m})},
\end{equation}
where $E_{0}$ is a constant, $\omega$ is the angular frequency, and $\mathbf{k}$ is the wave vector. The polarization vectors $\mathbf{e}_{ms}$ fulfill the transversality condition $\mathbf{k}\cdot\mathbf{e}_{ms}=0$ and the ortonormality condition $\mathbf{e}_{ms}^{\ast}\cdot\mathbf{e}_{ms^{\prime}}=\delta_{ss^{\prime}}$, while the annihilation operators $\hat{a}_{ms}$ satisfy the commutation relations $[\hat{a}_{ms},\hat{a}_{m^{\prime}s^{\prime}}^{\dagger}]=\delta_{mm^{\prime}}\delta_{ss^{\prime}}$ as well as $[\hat{a}_{ms},\hat{a}_{m^{\prime}s^{\prime}}]=0=[\hat{a}_{ms}^{\dagger},\hat{a}_{m^{\prime}s^{\prime}}^{\dagger}]$. After inserting Eq.~(\ref{Em}) into Eq.~(\ref{Gn}), and using the abbreviations $\kappa_{1}=m_{1}s_{1},\ldots,\kappa_{2n}=m_{2n}s_{2n}$, the $n$th-order correlation matrix for the slits $m_{1},\ldots,m_{2n}\in\{1,\ldots,L\}$ becomes \begin{subequations}
\begin{align}
\label{Gmodes}
&\mathbf{G}^{(n)}(x_{m_{1}},\ldots,x_{m_{2n}})=I_{0}^{n}\!\sum_{s_{1},\ldots,s_{2n}}W^{(n)}(\kappa_{1},\ldots,\kappa_{2n}) \nonumber \\
& \ \ \bigotimes_{\mu=1}^{n}\mathbf{e}_{\kappa_{\mu}}^{\dagger}
e^{-i(\mathbf{k}\cdot\mathbf{r}_{m_{\mu}}-\omega t_{m_{\mu}})}
\bigotimes_{\nu=n+1}^{2n}\mathbf{e}_{\kappa_{\nu}}
e^{i(\mathbf{k}\cdot\mathbf{r}_{m_{\nu}}-\omega t_{m_{\nu}})},
\end{align}
where we have defined $I_{0}=|E_{0}|^{2}$. The quantity
\begin{equation}
\label{W}
W^{(n)}(\kappa_{1},\ldots,\kappa_{2n})=\mathrm{tr}(\hat{\rho}\hat{a}_{\kappa_{1}}^{\dagger}\cdots\hat{a}_{\kappa_{n}}^{\dagger}
\hat{a}_{\kappa_{n+1}}\cdots\hat{a}_{\kappa_{2n}})
\end{equation}
is the $n$th-order cross-spectral density \cite{Mandelbook} that governs all the statistical properties of the photon field. \end{subequations}

We next expand the state density operator in Eq.~(\ref{W}) as $\hat{\rho}=\sum_{N=0}^{\infty}p_{N}\hat{\rho}^{(N)}$, in which $\hat{\rho}^{(N)}$ are $N$-photon states and $p_{N}$ are probability distributions. The average photon number of the whole state $\hat{\rho}$ is thereby $\bar{N}=\sum_{N=0}^{\infty}p_{N} N$. In atomic physics, various superselection rules exclusively admit density operators of such ``block-diagonal'' form, as superposition states of different number of particles (``off-diagonal'' terms) are prohibited within first quantization. Because Eq.~(\ref{W}) preserves the total number of photons, we may consider multiphoton states of the above form without loss of generality.

\begin{subequations}
In the first-quantized (particle) formalism, the rescaled matrix elements of the $N$-photon state $\hat{\rho}^{(N)}$ are given by
\begin{equation}
\label{elements}
\rho^{(N)}_{i_{1} \ldots i_{N},j_{1} \ldots j_{N}}\coloneqq N!\bra{i_{1},\ldots,i_{N}}\hat{\rho}^{(N)}\ket{j_{1},\ldots,j_{N}},
\end{equation}
where $i_{\alpha}$ and $j_{\alpha}$ with $\alpha\in\{1,\ldots,N\}$ denote individual photons in some modes. So if the number of modes is $K$, then $i_{1},\ldots,i_{N},j_{1},\ldots,j_{N}\in\{1,\ldots,K\}$. The correlations among $M\leq N$ photons in $\hat{\rho}^{(N)}$ are characterized by the \textit{reduced} $M$-photon density matrix $\varrho^{(N;M)}$. Its elements
\begin{align}
\label{reducedM}
&\varrho^{(N;M)}_{i_{1}\ldots i_{M},j_{1}\ldots j_{M}}\coloneqq \nonumber \\ 
&\frac{1}{(N-M)!}\sum_{l_{M+1},\ldots,l_N} \rho^{(N)}_{i_{1} \ldots i_{M}l_{M+1}\ldots l_N,j_{1} \ldots j_{M}l_{M+1}\ldots l_N}
\end{align}
fulfill the second-quantized (mode) identity \cite{footnote4}
\begin{equation}
\label{Red2}
\varrho^{(N;M)}_{i_{1}\ldots i_{M},j_{1}\ldots j_{M}}=\mathrm{tr}\big[\hat{\rho}^{(N)}\hat{a}_{j_{1}}^{\dagger}\cdots\hat{a}_{j_{M}}^{\dagger}
\hat{a}_{i_{1}}\cdots\hat{a}_{i_{M}}\big].
\end{equation}
Symmetry in $\hat{\rho}^{(N)}$ allows reduction of Eq.~(\ref{elements}) for any set of $N-M$ photons. Elements of the reduced $M$-photon density matrix $\varrho^{(M)}$ of the whole state $\hat{\rho}$ thus obey
\begin{equation}
\label{rhoM}
\varrho^{(M)}_{i_{1}\ldots i_{M},j_{1}\ldots j_{M}}=\sum_{N=0}^{\infty}p_{N}\varrho^{(N;M)}_{i_{1}\ldots i_{M},j_{1}\ldots j_{M}},
\end{equation}
being the full generalization of Eq.~(\ref{reduced2}). \end{subequations}

Combining Eqs.~(\ref{W}), (\ref{Red2}), and (\ref{rhoM}) now yields
\begin{equation}
\label{Wreduced}
W^{(n)}(\kappa_{1},\ldots,\kappa_{2n})=\varrho^{(n)}_{\kappa_{n+1}\ldots\kappa_{2n},\kappa_{1}\ldots\kappa_{n}}.
\end{equation}
Equation~(\ref{Wreduced}) establishes a fundamental relation between the field correlations and the particle correlations, stating that the $n$th-order cross-spectral density and thereby the $n$th-order correlation (coherence) matrix in Eq.~(\ref{Gmodes}) are fully specified by the reduced $n$-photon density matrix. We emphasize that even if first-quantized reduced density matrices are ubiquitous in atomic physics, they are rarely met in quantum optics.

\emph{First-order correlations and complementarity.}---Let us investigate the implications of Eq.~(\ref{Wreduced}) in the context of first-order correlations. For this we introduce the vector
\begin{equation}
\label{E}
\hat{\boldsymbol{\mathcal{E}}}=\big[\hat{E}_{1}^{(+)}\!(x_{1}), \hat{E}_{2}^{(+)}\!(x_{1}), \ldots, \hat{E}_{1}^{(+)}\!(x_{L}), \hat{E}_{2}^{(+)}\!(x_{L})\big]\!,
\end{equation}
where $\hat{E}_{s}^{(+)}(x_{m})$ is the $s\in\{1,2\}$ polarized part of the field operator in slit $m\in\{1,\ldots,L\}$, as given in Eq.~(\ref{Em}). We then construct the \textit{first-order field density matrix} \begin{subequations}
\begin{equation}
\label{G1}
\boldsymbol{\mathcal{G}}^{(1)}=\mathrm{tr}\big(\hat{\rho}\,\hat{\boldsymbol{\mathcal{E}}}^{\dagger}\!\otimes\hat{\boldsymbol{\mathcal{E}}}\big),
\end{equation}
which on using the correlation matrices in Eq.~(\ref{Gmodes}) can be expressed via the $2L \times 2L$ block representation
\begin{equation}
\label{G1block}
\boldsymbol{\mathcal{G}}^{(1)}=
\begin{bmatrix}
\mathbf{G}^{(1)}(x_{1},x_{1}) & \cdots & \mathbf{G}^{(1)}(x_{1},x_{L}) \\
\vdots & \ddots & \vdots \\
\mathbf{G}^{(1)}(x_{L},x_{1}) & \cdots & \mathbf{G}^{(1)}(x_{L},x_{L})
\end{bmatrix}
\!.
\end{equation}
Thus $\boldsymbol{\mathcal{G}}^{(1)}$ is Hermitian and contains all the information on the first-order field correlations in the configuration: the diagonal block terms are the polarization matrices in the slits, characterizing \textit{intra-field correlations}, while the off-diagonal block terms are the coherence matrices among the slits, describing \textit{inter-field correlations}. \end{subequations}\begin{subequations}

Analogously to polarimetric purity \cite{Gil07,Gil18}, we define the \textit{first-order degree of field purity} $F^{(1)}$ according to
\begin{equation}
\label{F1}
F^{(1)}=\frac{\mathcal{K}}{\mathcal{K}-1}\Big[\big\|\tilde{\boldsymbol{\mathcal{G}}}^{(1)}\big\|_{\mathrm{F}}^{2}-\frac{1}{\mathcal{K}}\Big]; \quad \tilde{\boldsymbol{\mathcal{G}}}^{(1)}=\frac{\boldsymbol{\mathcal{G}}^{(1)}}{\mathrm{tr}\boldsymbol{\mathcal{G}}^{(1)}},
\end{equation}
with $L\leq\mathcal{K}\leq2L$ being the number of modes considered (e.g., $\mathcal{K}=L$ if the field is fully polarized in all openings) and where the subscript F stands for the Frobenius norm. Because $\mathrm{tr}\mathbf{G}^{(1)}(x_{m},x_{m})=I_{0}\bar{N}_{m}$, with $\bar{N}_{m}$ being the average photon number in slit $m$, we have $\mathrm{tr}\boldsymbol{\mathcal{G}}^{(1)}=I_{0}\bar{N}$. The degree of purity obeys $0\leq F^{(1)}\leq1$ and is a measure of all the first-order field correlations within the system. The maximum $F^{(1)}=1$ is saturated when the field is completely polarized and first-order coherent at all slits. In this case the correlation matrices in Eq.~(\ref{G1block}) factorize in the respective space-time variables \cite{Glauber63,Mandelbook} and merely one eigenvalue of the field density matrix $\boldsymbol{\mathcal{G}}^{(1)}$ is nonzero. Likewise, the minimum $F^{(1)}=0$ takes place if the field is unpolarized and first-order incoherent for all openings, with additionally $\bar{N}_{1}=\ldots=\bar{N}_{L}$, and in such scenario all eigenvalues are equal.

The first-order particle correlations in the system are described by the one-photon reduced density matrix $\varrho^{(1)}$. To quantify the amount of such particle correlations, we utilize the linear entropy of the trace-normalized $\varrho^{(1)}$,
\begin{equation}
\label{S1}
S^{(1)}=\frac{\mathcal{K}}{\mathcal{K}-1}\big\{1-\mathrm{tr}\big[\tilde{\varrho}^{(1)}\big]^{2}\big\}; \quad \tilde{\varrho}^{(1)}=\frac{\varrho^{(1)}}{\mathrm{tr}\varrho^{(1)}},
\end{equation}
with $\mathrm{tr}\varrho^{(1)}=\bar{N}$ and the prefactor ensuring $0\leq S^{(1)}\leq1$. The upper bound $S^{(1)}=1$ stands for maximal first-order particle correlations in the \textit{full} $\bar{N}$-photon state $\hat{\rho}$ and is only met when $\tilde{\varrho}^{(1)}$ is maximally mixed. The lower bound $S^{(1)}=0$ corresponds to total lack of first-order particle correlations and is solely encountered if $\tilde{\varrho}^{(1)}$ is pure.

From Eqs.~(\ref{Gmodes}), (\ref{Wreduced}), and (\ref{G1block}) we learn that
\begin{equation}
\label{purity1}
\big\|\tilde{\boldsymbol{\mathcal{G}}}^{(1)}\big\|_{\mathrm{F}}^{2}=\mathrm{tr}\big[\tilde{\varrho}^{(1)}\big]^{2},
\end{equation}
and on further combining Eqs.~(\ref{F1})--(\ref{purity1}) we end up with the first  main result of this Letter:\end{subequations}
\begin{equation}
\label{F1S1}
F^{(1)}+S^{(1)}=1.
\end{equation}
Equation~(\ref{F1S1}) forms a fundamental quantum complementarity relation between the field correlations represented by $F^{(1)}$ and the particle correlations represented by $S^{(1)}$. It shows that these two quantum correlation species are \textit{mutually exclusive}, i.e., a variation of $F^{(1)}$ or $S^{(1)}$ alters its complementary partner such that the totality of first-order correlations within the system remains unchanged. Equation~(\ref{F1S1}) is very general as it sets no restrictions on the quantum state or the number of slits; it covers any mixed state in multislit vector-light interference.

\emph{Wave-particle triality.}---The arguably most recognized manifestion of complementarity is wave-particle duality, restricting the coexistence of ``which-path information'' and intensity-fringe visibility of quantum objects \cite{Greenberger88,Jaeger95,Englert96}. Photons, however, can exhibit interference not merely in terms of intensity fringes but also via polarization-state fringes, a unique characteristic of vector-light fields with no correspondence in scalar-light interference \cite{footnote3,Setala06a,Leppanen14}. Implications of such \textit{polarization modulation} in quantum-light complementarity have been recently studied in the celebrated double-slit configuration, leading to previously unexplored, fundamental physical findings about photon wave-particle duality \cite{Norrman17,Norrman18}.

\begin{subequations}
Let us examine vector-light photon interference in the conventional double-slit setup with one mode in each slit. The space-time points $x_{1}$ and $x_{2}$ correspond to the two slits in an opaque screen $\mathcal{A}$, while $x$ is the point at screen $\mathcal{B}$ where the light is observed within the paraxial regime. The \textit{intensity distinguishability} (or \textit{path predictability} at the single-photon level \cite{Greenberger88,Jaeger95,Englert96}) at $\mathcal{A}$ is defined as \cite{Norrman17}
\begin{equation}
\label{D}
D(x_{1},x_{2})=\frac{|\bar{N}_{1}-\bar{N}_{2}|}{\bar{N}_{1}+\bar{N}_{2}},
\end{equation}
in which $\bar{N}_{1}$ and $\bar{N}_{2}$ are the average photon numbers (or intensities) in openings 1 and 2, respectively, as before. The maximum $D(x_{1},x_{2})=1$ is saturated if all photons pass only via one slit, while the minimum $D(x_{1},x_{2})=0$ takes place when the intensities in the slits are equal.

In the observation plane $\mathcal{B}$, the photon number \textit{and} the polarization-state variations of the vector-light field are described by the \textit{total visibility} \cite{Norrman17}
\begin{equation}
\label{V}
V(x)=\frac{2(\bar{N}_{1}\bar{N}_{2})^{1/2}}{\bar{N}_{1}+\bar{N}_{2}}g^{(1)}(x_{1},x_{2}),
\end{equation}
including the \textit{degree of vector-light coherence}
\begin{equation}
\label{g}
g^{(1)}(x_{1},x_{2})=\frac{\|\mathbf{G}^{(1)}(x_{1},x_{2})\|_{\mathrm{F}}}{[\mathrm{tr}\mathbf{G}^{(1)}(x_{1},x_{1})\mathrm{tr}\mathbf{G}^{(1)}(x_{2},x_{2})]^{1/2}}.
\end{equation}
The degree of coherence satisfies $0\leq g^{(1)}(x_{1},x_{2})\leq1$ and thus also the total visibility is bounded as $0\leq V(x)\leq1$. Equation~(\ref{V}) is the vector-light generalization of the usual visibility relation met in the scalar framework that characterizes the variations of \textit{all four} Stokes parameters at $\mathcal{B}$ \cite{Norrman17}. Therefore, the total visibility does not generally coincide with the intensity visibility but it encompasses \textit{also} the polarization modulation. For instance, a photon in an even and orthogonally polarized superposition at the slits leads to $V(x)=1$, with maximal polarization-state fringes (due to first-order coherence), although in this case the intensity-fringe visibility is zero \cite{Norrman17}.
\end{subequations}\begin{subequations}

From Eqs.~(\ref{Gmodes}), (\ref{G1block}), (\ref{F1}), and (\ref{D})--(\ref{g}) we now obtain that in this double-slit vector-light scenario
\begin{equation}
\label{DV}
F^{(1)}=D^{2}(x_{1},x_{2})+V^{2}(x)\leq1,
\end{equation}
which is the recently reported complementarity relation for vectorial quantum light \cite{Norrman17}. Such complementarity, or wave-particle duality, is thereby already an inherent part and an important physical manifestation of $F^{(1)}$. The upper limit in Eq.~(\ref{DV}) is always saturated when the photon field is first-order coherent (in the vector sense), viz., $g(x_{1},x_{2})=1$. This holds for any pure single-photon vector-light state, yet not for pure quantum states in general \cite{Norrman17}, and highlights very essential physics about wave-particle duality in vector-light interference: for vectorial light the photon path predictability couples not to the intensity visibility (as for coherent scalar light in the classical domain \cite{Eberly17}) but instead to the total visibility which accounts also for the polarization modulation.

If the field is not first-order coherent, i.e., $g(x_{1},x_{2})<1$, then $D^{2}(x_{1},x_{2})+V^{2}(x)<1$ and thus there is no longer a strict complementarity between the path predictability and total visibility. Remarkably, however, on combining Eqs.~(\ref{F1S1}) and (\ref{DV}) we discover that
\begin{equation}
\label{triality}
D^{2}(x_{1},x_{2})+V^{2}(x)+S^{(1)}=1,
\end{equation}
stating that the particle correlations complement the lack of field correlations (and vice versa) such that the whole system is governed by a strict equality, a \textit{triality identity}. It should not be confused with triality relations involving the concurrence and intensity visibility \cite{Jakob10,Qian18}, since $S^{(1)}$ quantifies correlations among indistinguishable particles and $V(x)$ characterizes also the polarization modulation. Yet, as $S^{(1)}$ is specified by the reduced one-photon state, Eq.~(\ref{triality}) is in line with the standpoint that wave-particle duality/triality of light is a single-photon feature \cite{Qian18,Qian19}. Equation~(\ref{triality}) thus unveils a fundamental physical facet concerning vector-light quantum complementarity in the double-slit setup, and constitutes the second main result of this Letter. 

In fact, Eq.~(\ref{triality}) encompasses three specific dualities:
\begin{align}
\label{duality1}
D^{2}(x_{1},x_{2})+V^{2}(x)&=1, \quad \mathrm{if} \  g(x_{1},x_{2})=1, \\
\label{duality2}
D^{2}(x_{1},x_{2})+S^{(1)}&=1, \quad \mathrm{if} \  g(x_{1},x_{2})=0, \\
\label{duality3}
V^{2}(x)+S^{(1)}&=1, \quad \mathrm{if} \ \bar{N}_{1}=\bar{N}_{2}.  \ \ \ \ \; \;
\end{align}
The first case, Eq.~(\ref{duality1}), is the usual scenario with first-order coherent fields and no first-order particle correlations; all pure single-photon states fall into this category. The second case, Eq.~(\ref{duality2}), is encountered for fields being first-order incoherent and thus displaying no intensity or polarization-state variations in the observation plane; yet first-order correlations among particles are possible. The last case, Eq.~(\ref{duality3}), is met for zero path predictability and in this scenario both first-order field correlations as well as first-order particle correlations can be present. For other cases, $0<g^{(1)}(x_{1},x_{2})<1$ and $\bar{N}_{1}\neq\bar{N}_{2}$, the system is governed by the triality relation (\ref{triality}).
\end{subequations}

\emph{Higher-order correlations and complementarity.}---As a final point we investigate quantum complementarity with higher-order correlations. To this end we use $\hat{\boldsymbol{\mathcal{E}}}$ in Eq.~(\ref{E}) and introduce the general \textit{$n$th-order field density matrix}
\begin{equation}
\label{calGn}
\boldsymbol{\mathcal{G}}^{(n)}=\mathrm{tr}\big[\hat{\rho}\,(\hat{\boldsymbol{\mathcal{E}}}^{\dagger})^{\otimes n}
\otimes(\hat{\boldsymbol{\mathcal{E}}})^{\otimes n}\big],
\end{equation}
which is Hermitian, non-negative definite, and covers all possible $n$th-order intra-field and inter-field correlations. The \textit{$n$th-order degree of field purity} $F^{(n)}$ is defined as \begin{subequations}
\begin{equation}
\label{Fn}
F^{(n)}=\frac{\mathcal{K}^{n}}{\mathcal{K}^{n}-1}\Big[\big\|\tilde{\boldsymbol{\mathcal{G}}}^{(n)}\big\|_{\mathrm{F}}^{2}-\frac{1}{\mathcal{K}^{n}}\Big]; \ \ \tilde{\boldsymbol{\mathcal{G}}}^{(n)}=\frac{\boldsymbol{\mathcal{G}}^{(n)}}{\mathrm{tr}\boldsymbol{\mathcal{G}}^{(n)}},
\end{equation}
which satisfies the constraint $0\leq F^{(n)}\leq1$ for all orders. When $F^{(n)}=1$ the vector-light field is completely $n$th-order coherent \cite{Glauber63,Mandelbook}.

The information on the $n$th-order particle correlations is encoded in the reduced $n$-photon density matrix $\varrho^{(n)}$. Similarly to Eq.~(\ref{S1}), we quantify the amount of these correlations in the \textit{whole} state $\hat{\rho}$ via the linear entropy
\begin{equation}
\label{Sn}
S^{(n)}=\frac{\mathcal{K}^{n}}{\mathcal{K}^{n}-1}\big\{1-\mathrm{tr}\big[\tilde{\varrho}^{(n)}\big]^{2}\big\}; \quad \tilde{\varrho}^{(n)}=\frac{\varrho^{(n)}}{\mathrm{tr}\varrho^{(n)}},
\end{equation}
with the normalizations ensuring $0\leq S^{(n)}\leq1$ for all $n$. The bound $S^{(n)}=1$ [$S^{(n)}=0$] is only met when $\tilde{\varrho}^{(n)}$ is maximally mixed (pure), representing maximal (full lack of) $n$th-order particle correlations in the system.

Equations~(\ref{Gmodes}), (\ref{Wreduced}), (\ref{E}), and (\ref{calGn}) now yield 
\begin{equation}
\label{purityn}
\big\|\tilde{\boldsymbol{\mathcal{G}}}^{(n)}\big\|_{\mathrm{F}}^{2}=\mathrm{tr}\big[\tilde{\varrho}^{(n)}\big]^{2},
\end{equation}
which together with Eqs.~(\ref{Fn}) and (\ref{Sn}) eventually leads to the third and last major discovery of this Letter:\end{subequations}
\begin{equation}
\label{FnSn}
F^{(n)}+S^{(n)}=1.
\end{equation}
Equation~(\ref{FnSn}) states that there is a \textit{strict trade-off} among field correlations and particle correlations for any $n$, such that the total sum of quantum correlations is invariant. The first-order relation (\ref{F1S1}) and triality identity (\ref{triality}) follow from this fundamental complementarity relation. We may thus view Eq.~(\ref{FnSn}) as the most general statement of field-particle duality as regards quantum correlations in vector-light interference; it covers any number of slits, any correlation orders, and any states of light.

\emph{Conclusions.}---In summary, we have studied quantum correlations in general multislit vector-light interference and established a fundamental complementarity relation between the field correlations and particle correlations that holds for any correlation orders and quantum states. It dictates that these two quantum correlation species are not interchangeable but mutually exclusive and reflects the intrinsic field-particle nature of light at a deep level. For this, we derived a link which connects the $n$th-order field-coherence matrix and the reduced $n$-photon state. The relation was also used to study photon wave-particle duality in the double-slit setup, showing that the particle correlations complement the vector-light duality inequality by transforming it into an equality, a triality identity. Our work thus provides a quantum information-theoretic foundation of optical coherence for vector-light fields, uncovers fundamental aspects concerning field-particle complementarity, and identifies directions towards future research involving quantum interference.

\emph{Acknowledgments.}---We thank A. T. Friberg, A. Z. Khoury, S. Franke-Arnold, F. De Zela, K. Rz\k{a}\.zewski, K. Paw{\l}owski, and R. Seiringer for fruitful discussions and correspondence. A. Norrman acknowledges the Swedish Cultural Foundation in Finland for financial support.


\begin{thebibliography}{13}
\bibitem{footnote2} In atomic physics, one usually defines reduced one-boson states of $N$ bosons with a prefactor of $N$. Consequently, reduced $M$-boson states of $N$ bosons get a prefactor of $N!/(N-M)!$, being the number of all sets of $M$ bosons selected out of $N$ bosons. While the full density matrix of $N$ bosons does not typically acquire a similar prefactor, for consistency we multiply it by $N!$ [see also Eqs.~(\ref{elements}) and (\ref{reducedM})]. In this way the density matrix of $N$ bosons can  be identified with its $N$-boson trivial reduction.
\end{thebibliography}
\end{document}